\documentclass[12pt,preprint]{aastex}
\usepackage{color}
\def\j0927{J0927+2943}
\def\het{{HET}}
\def\sdss{{SDSS}}

\def\hst{{HST}}

\def\etal{~et al.}
\def\simlt{\lower.5ex\hbox{$\; \buildrel < \over \sim \;$}}
\def\simgt{\lower.5ex\hbox{$\; \buildrel > \over \sim \;$}}

\def\gsim{\lower 2pt \hbox{$\, \buildrel {\scriptstyle >}\over
{\scriptstyle \sim}\,$}}
\def\lsim{\lower 2pt \hbox{$\, \buildrel {\scriptstyle <}\over
{\scriptstyle \sim}\,$}}

\def\deg{\ifmmode ^{\circ}
         \else $^{\circ}$\fi}
\def\pdeg{\ifmmode
           $\setbox0=\hbox{$^{\circ}$}\rlap{\hskip.11\wd0 .}$^{\circ}
     \else \setbox0=\hbox{$^{\circ}$}\rlap{\hskip.11\wd0 .}$^{\circ}$\fi}
     
\def\pc{\ifmmode \mathrm{pc} \else $\mathrm{pc}$ \fi}
\def\mpc{\ifmmode \mathrm{Mpc} \else $\mathrm{Mpc}$\fi}
\def\mpcthree{\ifmmode \mathrm{Mpc}^{-3} \else $\mathrm{Mpc}^{-3}$\fi}
\def\gpcthree{\ifmmode \mathrm{Gpc}^{-3} \else $\mathrm{Gpc}^{-3}$\fi}

\def\kelvin{\ifmmode \mathrm{K} \else {$\mathrm{K}$}\fi}
\def\kev{\ifmmode \mathrm{keV} \else $\mathrm{keV}$ \fi}

\def\lsun{\ifmmode {L_\odot} \else $L_\odot$\fi}
\def\msun{\ifmmode M_\odot \else $M_\odot$\fi}
\def\msunyr{\ifmmode M_\odot~\mathrm{yr}^{-1} \else $M_\odot~\mathrm{yr}^{-1}$\fi}

\def\cosi{\ifmmode {\cos\,i} \else $\cos\,i$\fi}

\def\heii{\ifmmode {\rm He{\sc ii}} \else He~{\sc ii}\fi}
\def\mgii{\ifmmode {\rm Mg{\sc ii}} \else Mg~{\sc ii}\fi}
\def\caii{\ifmmode {\rm Ca{\sc ii}} \else Ca~{\sc ii}\fi}
\def\ciii{\ifmmode {\rm C{\sc iii}]} \else C~{\sc iii}]\fi}
\def\civ{\ifmmode {\rm C{\sc iv}} \else C~{\sc iv}\fi}
\def\mgii{\ifmmode {\rm Mg{\sc ii}} \else Mg~{\sc ii}\fi}
\newcommand{\oiii}{{\sc [O~iii]}}
\newcommand{\oii}{{\sc [O~ii]}}

\newcommand{\neiii}{{[Ne~{\sc iii}]}}
\newcommand{\nev}{{[Ne~{\sc v}]}}
\newcommand{\sii}{{\sc [S~ii]}}

\newcommand{\fevii}{Fe~{\sc vii}}

\def\teff{\ifmmode {T_{\rm eff}} \else $T_{\rm eff}$\fi}
\def\tmax{\ifmmode {T_{\rm max}} \else $T_{\rm max}$\fi}

\def\mbh{\ifmmode {M_{\rm BH}} \else $M_{\rm BH}$\fi}
\def\led{\ifmmode L_{\mathrm{Ed}} \else $L_{\mathrm{Ed}}$\fi}
\def\lbolflare{\ifmmode L_{\mathrm{bol,flare}} \else $L_{\mathrm{bol,flare}}$\fi}
\def\lagn{\ifmmode L_{\mathrm{agn}} \else $L_{\mathrm{agn}}$\fi}
\def\lbolagn{\ifmmode L_{\mathrm{bol,agn}} \else $L_{\mathrm{bol,agn}}$\fi}
\def\lbol{\ifmmode L_{\mathrm{bol}} \else $L_{\mathrm{bol}}$\fi}
\def\mdot{\ifmmode {\dot M} \else $\dot M$\fi}
\def\mdoto{\ifmmode {\dot{M}_0} \else  $\dot{M}_0$\fi}
\def\mdotf{\ifmmode {\dot{M}_\mathrm{flare}} \else  $\dot{M}_\mathrm{flare}$\fi}

\def\hnot{\ifmmode H_0 \else H$_0$ \fi}

\def\vkep{\ifmmode v_\mathrm{Kep} \else $v_\mathrm{Kep}$ \fi}
\def\vc{\ifmmode v_\mathrm{c} \else $v_\mathrm{c}$ \fi}

\def\vthree{\ifmmode v_{1000} \else $v_{1000}$ \fi}
\def\vrel{\ifmmode v_\mathrm{rel} \else $v_\mathrm{rel}$ \fi}
\def\vkick{\ifmmode v_\mathrm{kick} \else $v_\mathrm{kick}$ \fi}
\def\vkickz{\ifmmode v_{\mathrm{kick},z} \else $v_{\mathrm{kick},z} $ \fi}
\def\vkicky{\ifmmode v_{\mathrm{kick},y} \else $v_{\mathrm{kick},y} $ \fi}
\def\vchar{\ifmmode v_\mathrm{char} \else $v_\mathrm{char}$ \fi}
\def\eflare{\ifmmode E_\mathrm{flare} \else $E_\mathrm{flare}$ \fi}
\def\ekick{\ifmmode E_\mathrm{kick} \else $E_\mathrm{kick}$ \fi}
\def\ecoll{\ifmmode E_\mathrm{coll} \else $E_\mathrm{coll}$ \fi}
\def\ezero{\ifmmode E_\mathrm{0} \else $E_\mathrm{0}$ \fi}
\def\efac{\ifmmode \xi_\mathrm{E} \else $\xi_\mathrm{E}$ \fi}
\def\tqso{\ifmmode t_\mathrm{QSO} \else $t_\mathrm{QSO}$ \fi}
\def\tflare{\ifmmode t_\mathrm{flare} \else $t_\mathrm{flare}$ \fi}
\def\tzero{\ifmmode t_\mathrm{0} \else $t_\mathrm{0}$ \fi}
\def\tfac{\ifmmode \xi_\mathrm{t} \else $\xi_\mathrm{t}$ \fi}
\def\gfac{\ifmmode f_\mathrm{g} \else $f_\mathrm{g}$ \fi}
\def\lflare{\ifmmode L_\mathrm{flare} \else $L_\mathrm{flare}$ \fi}
\def\fflare{\ifmmode F_\mathrm{flare} \else $F_\mathrm{flare}$ \fi}
\def\nflare{\ifmmode N_\mathrm{flare} \else $N_\mathrm{flare}$ \fi}
\def\tshock{\ifmmode T_\mathrm{shock} \else $T_\mathrm{shock}$ \fi}
\def\rmin{\ifmmode R_\mathrm{1} \else $R_\mathrm{1}$ \fi}
\def\rmax{\ifmmode R_\mathrm{2} \else $R_\mathrm{2}$ \fi}
\def\rbound{\ifmmode R_\mathrm{b} \else $R_\mathrm{b}$ \fi}
\def\pbound{\ifmmode P_\mathrm{b} \else $P_\mathrm{b}$ \fi}
\def\mbound{\ifmmode M_\mathrm{b} \else $M_\mathrm{b}$ \fi}
\def\mbo{\ifmmode M_{\mathrm{b}0} \else $M_{\mathrm{b}0} $ \fi}
\def\ebo{\ifmmode E_{\mathrm{b}0} \else $E_{\mathrm{b}0} $ \fi}
\def\efinal{\ifmmode E_\mathrm{final} \else $E_\mathrm{final} $ \fi}
\def\tbound{\ifmmode t_\mathrm{b} \else $t_\mathrm{b}$ \fi}
\def\tagn{\ifmmode t_\mathrm{AGN} \else $t_\mathrm{AGN}$ \fi}
\def\rlim{\ifmmode R_\mathrm{lim} \else $R_\mathrm{lim}$ \fi}
\def\vlim{\ifmmode v_\mathrm{lim} \else $v_\mathrm{lim}$ \fi}
\def\vphi{\ifmmode v_\phi \else $v_\phi$ \fi}
\def\mlim{\ifmmode M_\mathrm{lim} \else $M_\mathrm{lim}$ \fi}
\def\tlim{\ifmmode t_\mathrm{lim} \else $t_\mathrm{lim}$ \fi}
\def\llim{\ifmmode L_\mathrm{lim} \else $L_\mathrm{lim}$ \fi}
\def\fqso{\ifmmode f_\mathrm{QSO} \else $f_\mathrm{QSO}$ \fi}

\def\hbeta{\ifmmode \rm{H}\beta \else H$\beta$\fi}
\def\hbetan{\ifmmode \rm{H}\beta_{\rm n} \else H$\beta_{\rm n}$\fi}
\def\hgamma{\ifmmode \rm{H}\gamma \else H$\gamma$\fi}
\def\hdelta{\ifmmode \rm{H}\delta \else H$\delta$\fi}
\def\hepsilon{\ifmmode \rm{H}\epsilon \else H$\epsilon$\fi}
\def\hzeta{\ifmmode \rm{H}\zeta \else H$\zeta$\fi}
\def\halpha{\ifmmode \rm{H}\alpha \else H$\alpha$\fi}
\def\lalpha{\ifmmode \rm{Ly}\alpha \else Ly$\alpha$}

\def\dvhb{\ifmmode \Delta v_{\hbeta} \else $\Delta v_{\hbeta}$\fi}
\def\dvmg{\ifmmode \Delta v_{\rm{Mg}} \else $\Delta v_{\rm{Mg}}$\fi}

\def\muobs{\ifmmode {\mu_{o}} \else  $\mu_{o}$ \fi}
\def\cosi{\ifmmode {\mathrm{cos}\,i} \else $\mathrm{cos}\,i$\fi}

\def\teff{\ifmmode {T_{eff}} \else $T_{eff}$ \fi}
\def\tmax{\ifmmode {T_{max}} \else $T_{max}$ \fi}

\def\tauh{\ifmmode {\tau_{\rm H}} \else $\tau_{\rm H}$ \fi}

\def\yr{\ifmmode {\rm yr} \else  yr \fi}
\def\kms{\ifmmode \rm km~s^{-1}\else $\rm km~s^{-1}$\fi}
\def\cm{\ifmmode {\rm cm} \else  cm \fi}
\def\cmmitwo{\ifmmode \rm cm^{-2} \else $\rm cm^{-2}$\fi}
\def\cmmithree{\ifmmode \rm cm^{-3} \else $\rm cm^{-3}$\fi}
\def\cmps{\ifmmode \rm cm~s^{-1}\else $\rm cm~s^{-1}$\fi}
\def\cmpsps{\ifmmode \rm cm~s^{-2}\else $\rm cm~s^{-2}$\fi}
\def\kmps{\ifmmode \rm km~s^{-1}\else $\rm km~s^{-1}$\fi}
\def\kmpspmpc{\ifmmode \rm km~s^{-1}~Mpc^{-1} \else
    $\rm km~s^{-1}~Mpc^{-1}$\fi}
  
\def\gcmthree{\ifmmode \rm g~cm^{-3} \else $\rm g~cm^{-3}$\fi}
\def\gcmtwo{\ifmmode \rm g~cm^{-2} \else $\rm g~cm^{-2}$\fi}
   
\def\erg{\ifmmode {\rm erg} \else $\rm erg$ \fi}
\def\ergps{\ifmmode {\rm erg~s^{-1}} \else $\rm erg~s^{-1}$ \fi}
\def\ergcms{\ifmmode \rm erg~cm^{-2}~s^{-1} \else $\rm erg~cm^{-2}~s^{-1}$ \fi}
\def\ergcmshz{\ifmmode \rm erg~s^{-1}~cm^{-2}~Hz^{-1} \else $\rm
erg~cm^{-2}~s^{-1}~Hz^{-1}$ \fi}
\def\ergcmsa{\ifmmode \rm erg~cm^{-2}~s^{-1}~\AA^{-1} \else $\rm
erg~cm^{-2}~s^{-1}~\AA^{-1}$ \fi}
\def\ergshz{\ifmmode \rm erg s^{-1} Hz^{-1} \else
   $\rm erg s^{-1} Hz^{-1}$ \fi}

\def\lam{\ifmmode {\lambda} \else {$\lambda$} \fi}
\def\llam{\ifmmode {L_\lambda} \else  $L_\lambda$ \fi}
\def\lamLlam{\ifmmode \lambda L_{\lambda}(5100) \else {$\lambda L_{\lambda}(5100)$} \fi}
\def\nuLnu{\ifmmode \nu L_{\nu}(5100) \else {$\nu L_{\nu}(5100)$} \fi}
\def\ilam{\ifmmode {I_\lambda} \else  $I_\lambda$ \fi}
\def\flam{\ifmmode {F_\lambda} \else  $F_\lambda$ \fi}
\def\inu{\ifmmode {I_\nu} \else  $I_\nu$ \fi}
\def\fnu{\ifmmode {F_\nu} \else  $F_\nu$ \fi}
\def\bnu{\ifmmode {B_\nu} \else  $B_\nu$ \fi}

\def\msigma{\ifmmode M_{\sigma} \else $M_{\sigma}$\fi}
\def\mbulge{\ifmmode M_{\mathrm{bulge}} \else $M_{\mathrm{bulge}}$\fi}
\def\mgal{\ifmmode M_{\mathrm{gal}} \else $M_{\mathrm{gal}}$\fi}
\def\lgal{\ifmmode L_{\mathrm{gal}} \else $L_{\mathrm{gal}}$\fi}
\def\lbulge{\ifmmode L_{\mathrm{bulge}} \else $L_{\mathrm{bulge}}$\fi}
\def\mgalstar{\ifmmode M^*_{\mathrm{gal}} \else $M^*_{\mathrm{gal}}$\fi}

\def\mbhsigstar{\ifmmode M_{\mathrm{BH}} - \sigma_* \else $M_{\mathrm{BH}} - \sigma_*$ \fi}
\def\deltalogmbh{\ifmmode \Delta~{\mathrm{log}}~M_{\mathrm{BH}} \else $\Delta$~log~$M_{\mathrm{BH}}$\fi}

\def\sigstar{\ifmmode \sigma_* \else $\sigma_*$\fi}
\def\sigthree{\ifmmode \sigma_{\mathrm{[O~III]}} \else $\sigma_{\mathrm{[O~III]}}$\fi}
\def\sigtwo{\ifmmode \sigma_{\mathrm{[O~II]}} \else $\sigma_{\mathrm{[O~II]}}$\fi}
\def\signl{\ifmmode \sigma_{\mathrm{NL}} \else $\sigma_{\mathrm{NL}}$\fi}
\def\wthree{\ifmmode {\rm FWHM({[O~III]})} \else $FWHM({[O~III]})$ \fi}
\def\wtwo{\ifmmode {\rm FWHM({[O~II]})} \else $FWHM({[O~II]})$ \fi}
\def\mthree{\ifmmode M_{\mathrm [O~III]} \else $M_{\mathrm [O~III]}$ \fi}
\def\mtwo{\ifmmode M_{\mathrm [O II]} \else $M_{\mathrm [O II]}$ \fi}

\def\lbreak{\ifmmode L_{\mathrm{break}} \else $L_{\mathrm{break}}$\fi}
\def\lcut{\ifmmode L_{\mathrm{cut}} \else $L_{\mathrm{cut}}$\fi}

\received{}
\accepted{}
\journalid{}{}
\articleid{}{}

\slugcomment{Submitted to ApJ}

\shortauthors{Shields, Bonning, \& Salviander}
\shorttitle{Recoiling Quasar J0927+2943}

\begin{document}

\title{Comment on the Black Hole Recoil Candidate Quasar SDSS~J092712.65+294344.0}

\author{G. A. Shields\altaffilmark{1}, E.~W. Bonning \altaffilmark{1,2}, S.~Salviander \altaffilmark{1}}

\altaffiltext{1}{Department of Astronomy, University of Texas, Austin,
TX 78712; shields@astro.as.utexas.edu} 

\altaffiltext{2}{YCAA - Department of Physics, Yale University, New Haven, CT 06520; erin.bonning@yale.edu}

\begin{abstract}

SDSS J092712.65+294344.0 has been proposed as a candidate for a supermassive black hole ($\sim 10^{8.8}~\msun$) ejected at high speed from the host galactic nucleus by gravitational radiation recoil,
or alternatively for a supermassive black hole binary.  This is based on
a blueshift of 2650 \kms\ of the broad emission lines (``b-system'') relative to the narrow emission lines
(``r-system'') presumed to reflect the galaxy velocity.  New observations with the Hobby-Eberly Telescope (\het) confirm the essential features of the spectrum.   We note a third redshift system,
characterized by  weak,
narrow emission lines  of \oiii\ and \oii\  at an intermediate velocity $900~\kms$ redward of the broad line velocity (``i-system'').   A composite spectrum of SDSS QSOs similar to \j0927\ illustrates
the feasibility of detecting the calcium K  absorption line in spectra of sufficient quality. 
The i-system may represent the QSO host galaxy or a companion.
Photoionization  requires the black hole to be $\sim3$~kpc  from the r-system emitting gas,
implying that we are observing the system only $10^6~\yr$ after the recoil event and contributing to the
low probability of observing such a system.   The \het\ observations give an upper limit of
$10~\kms$ per year on the rate of change of the velocity difference between the r- and b-systems,
constraining the orbital phase in the binary model.  These considerations and the
presence of a cluster of galaxies apparently containing \j0927\ favor the idea that this system represents
a superposition of two AGN.

\end{abstract}


\section{Introduction}
\label{sec:intro}

Simulations of binary black hole mergers  show large recoil velocities
(``kicks'') of the final merged black hole resulting from
asymmetric emission of gravitational radiation.
\cite{campa07b} predict a maximum recoil velocity of $4000~\kmps$ for
equal mass black holes with maximal spin ($a_*=1$),
with the spins anti-aligned and lying in the orbital plane.
Kicks as large as 3300~\kms\ have been found in numerical simulations \citep{dain08}
for $a_* = 0.92$.
For $a_* = 0.9$, Baker et al (2008) predict that as many as 
one-quarter of mergers will give kicks over
$1000~\kms$ for random spin orientations and 
mass ratios distributed in the range 0.25 to 1.  For  a binary supermassive
black hole ($\sim10^8~M_\odot$) formed during a galactic merger
\citep{begelman80}, the kick may displace the black hole from the
galactic nucleus or eject it entirely \citep[][and  references
therein]{merritt04}.  For a recoil occurring in an active galactic
nucleus (AGN) with an accretion disk, the inner disk will remain bound to
the black hole, providing fuel for continued AGN activity \citep{loeb07}.   
This might be observed as a QSO  displaced  from the
galactic nucleus \citep{madau04, loeb07}, as a QSO with emission lines shifted relative to
the galactic velocity \citep{bonning07}, as thermal emission from shocked gas in the disk
\citep{lippai08,shields08,schnitt08}, or as flares from tidal disruption of stars bound
to the moving hole \citep{komerritt08}.
However,  AGN rarely show displaced nuclei
\citep{libeskind06}, and \citet{bogda07} argue that accretion by the
merging black holes will align their spins in a way unfavorable for
large kicks.  

\citet{bonning07} conducted a search for recoil candidates among QSOs in the
Sloan Digital Sky Survey (SDSS).  They looked for cases in which the velocity of the broad emission lines differed substantially from that of the narrow lines.  The inner disk and broad line region (BLR) should remain bound to the black hole, the narrow line region (NLR) will be left behind.  Photoionization of the residual nuclear gas and the general interstellar medium by the displaced QSO should give narrow emission lines at the velocity of the host galaxy.  \citet{bonning07} found many cases of shifted broad lines (as was previously known) up to $2600~\kms$, but they attributed most cases to physical processes in the BLR rather than recoil.  They listed several spectroscopic criteria to be satisfied by viable candidates for true kicks, including
symmetrical broad line profiles, agreement of the velocity of the broad \hbeta\ and \mgii\ lines, and agreement of the profiles of the high and low ionization narrow lines.  \citet{bonning07} listed two candidate objects that satisfied these criteria, with velocity shifts of $\sim500~\kms$.

\begin{figure}[ht]
\begin{center}
\plotone{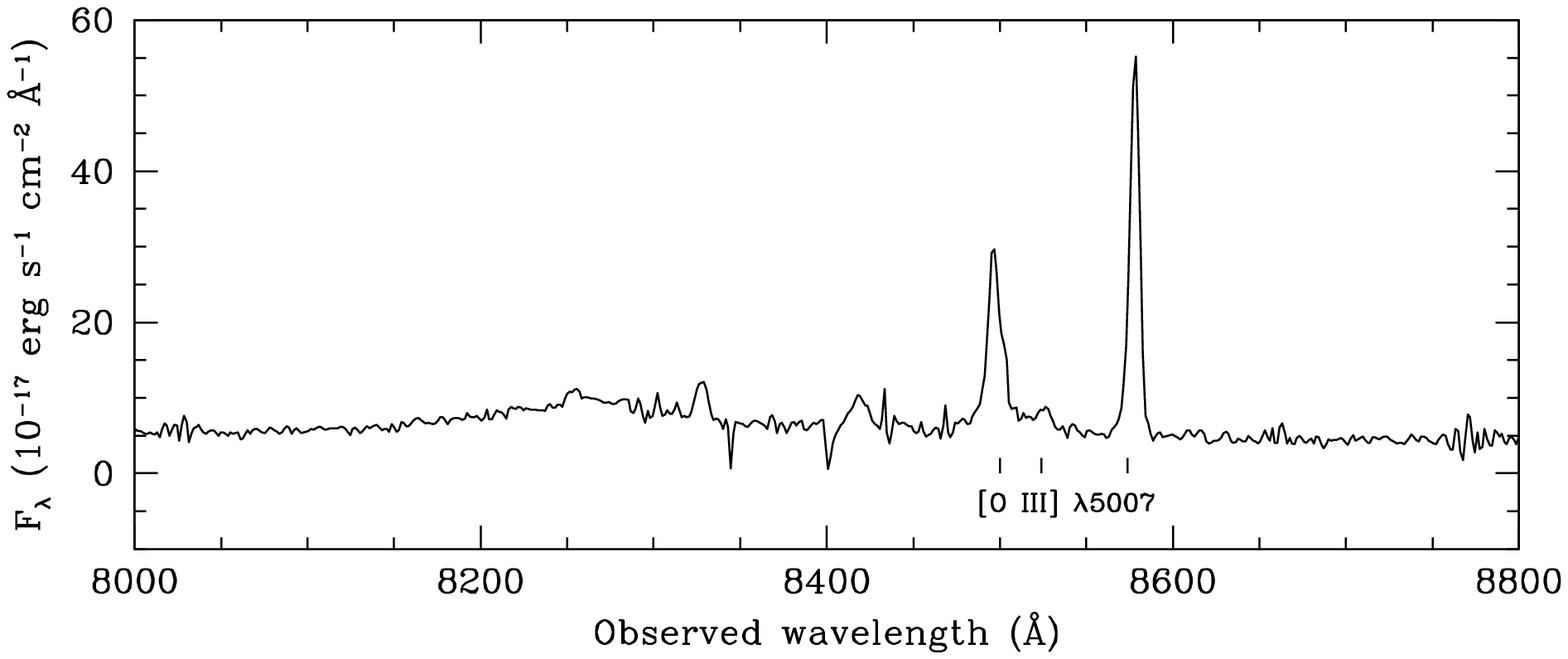}
\figcaption[]{
SDSS spectrum of \j0927 in the \hbeta\ region.   Note \oiii~$\lambda5007$ emission line in the
r-system  at $\lambda8578$ and in the proposed i-system at $\lambda8526$.
The emission feature at $\lambda8500$ is a blend of \oiii~$\lambda4959$ in the r-system
and $\lambda5007$ in the b-system.
See text for discussion.
\label{fig:oiiifig} }
\end{center}
\end{figure}

\begin{figure}[ht]
\begin{center}
\plotone{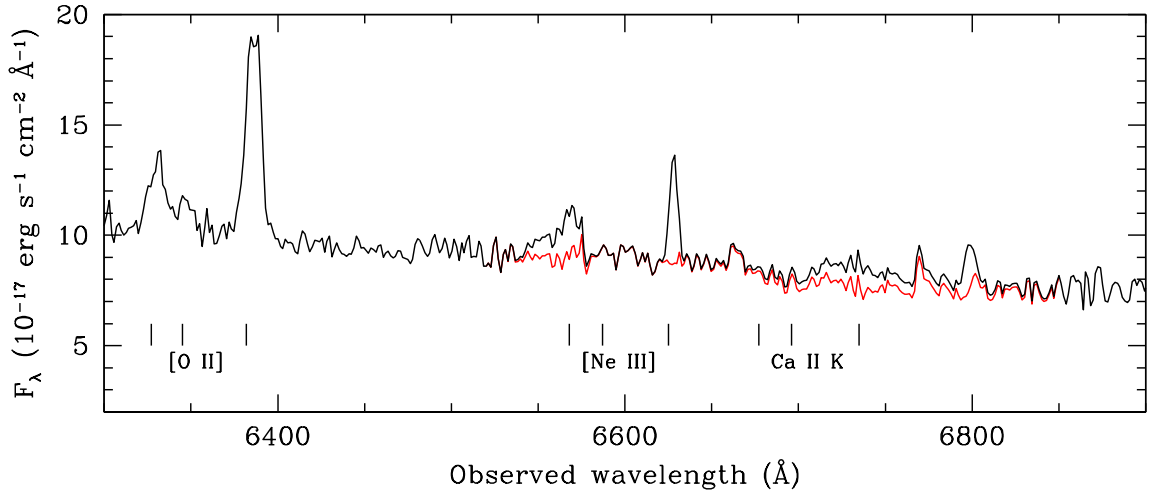}
\figcaption[]{
SDSS spectrum of \j0927
in the  region of the stellar calcium H and K lines with (red) and without subtraction of
\neiii~$\lambda3968$ and \hepsilon\ at the b- and r-system redshifts.  Indicated are the
wavelengths of \oii~$\lambda3727$, \neiii~$\lambda3968$, and  \caii~$\lambda3933$
at the three redshifts.  Note the weak i-system \oii\ emission line.
See text for discussion.
\label{fig:caiifig}
 }
\end{center}
\end{figure}

\citet{komossa08} proposed the QSO SDSS J092712.65+294344.0 as a promising candidate for black hole recoil.   
The object shows broad emission lines of \hbeta\   and
\mgii\ at $z_b = 0.6977$ (b-system) and narrow emission lines at $z_r = 0.7128$ (r-system). 
Portions of the spectrum are shown  in Figures \ref{fig:oiiifig} and \ref{fig:caiifig};  see
\citet{komossa08} for the full \sdss\ spectrum with line identifications.  The
\hbeta\ and \mgii\ lines have symmetrical profiles with closely consistent blueshifts of $2650~\kms$ relative to the narrow lines.   \citet{komossa08} identify the narrow lines with the host galaxy and the broad lines with a recoiling black hole ejected from the nucleus with a line-of-sight velocity of
$2650~\kms$ toward the observer.  The narrow lines have normal AGN line intensity ratios, with unusually narrow profiles ($\mathrm{FWHM} = 170~\kms$).  There is no indication of broad lines at the redshift of the narrow lines.  Unusual for QSOs with shifted broad lines, there is an additional set of narrow emission lines at the peak velocity of the broad lines.  \citet{komossa08} suggest that the narrowness of the r-system line profiles is consistent with photoionization of gas in a galactic disk by an ionizing source displaced from the nucleus, and that
the b-system narrow lines may represent weakly bound gas moving with the black hole.  They further argue that a chance superposition of two AGN with these properties is unlikely, and that the velocity difference is too large for a merger of two active galaxies in a cluster of galaxies.  
 More recently, \citet{bogda08} and \citet{dotti08} have proposed the alternative hypothesis that
 \j0927\ represents a supermassive binary black hole system whose orbital velocity leads to
 the velocity of the b-system while the r-system represents the host galaxy.

Here  we present a new \het\ spectrum of
\j0927 that confirms the basic features of the system and sets limits on
variability.  We note the presence of a third redshift system in \j0927, which is closer to the broad line velocity and has implications for the interpretation of the system.  We emphasize the importance of identifying the stellar absorption lines and discuss their possible presence at the intermediate redshift.  We also consider photoionization constraints on the geometry and age of the system on the recoil hypothesis. Finally. we comment on the relative merits of the recoil, binary, and superposition hypotheses and potential
tests.

\section{HET spectrum}
\label{sec:hetspec}

\begin{figure}[ht]
\begin{center}
\plotone{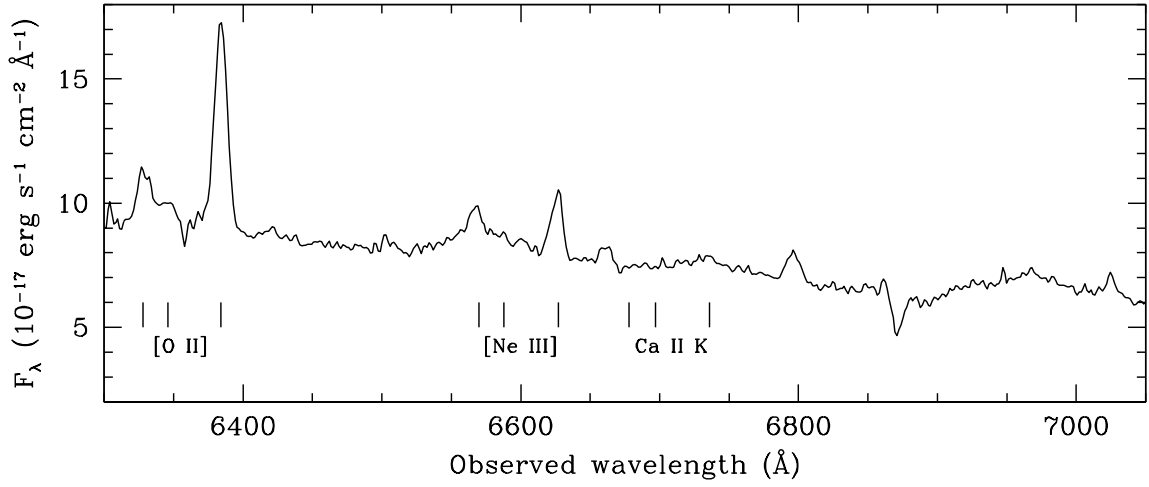}
\figcaption[]{Optical spectrum of \j0927\ obtained with the Hobby Eberly Telescope at
McDonald Observatory.  Indicated are the
wavelengths of \oii~$\lambda3727$, \neiii~$\lambda3968$, and  \caii~$\lambda3933$
at the three redshifts.    The dip at 6870~\AA\ is an artifact. 
\label{fig:hetfig} }
\end{center}
\end{figure}

In order to confirm the basic features of the spectrum and to search for possible stellar
absorption lines (see below), we observed \j0927 using the Hobby-Eberly Telescope (HET) at
McDonald Observatory.  Spectra were obtained with the Low Resolution
Spectrograph (LRS) with the G2 grism and a 2~arcsec slit giving a spectral resolution
of $\mathrm{FWHM} = 410~\kms$ as measured from the night sky lines.   Integrations of
40 minutes were obtained on the evenings of May 8 and May 9, 2008 (civil),
each divided into two cosmic ray splits.  The data were reduced using standard
procedures.   The relative flux calibration as a function of wavelength should be accurate,
but absolute calibration is not precise because of the nature of the \het, which has a changing
effective aperture during the integration.  The flux-calibrated spectrum is shown in Figure \ref{fig:hetfig}.
The spectrum covers the observed wavelength range 4300 to 7250~\AA, which includes \mgii\ through \hgamma\ but not  the \hbeta-\oiii\ region.  The \het\ spectrum confirms the basic features
of the SDSS spectrum, including the r- and b-redshift systems representing the strong
narrow lines and the broad lines, respectively.
The equivalent widths for the \het\ and SDSS spectra, measured with a Gaussian profile
with the IRAF task SPLOT \footnote{IRAF is distributed by the National Optical Astronomy Observatories, which are operated by the Association of Universities for Research in Astronomy, Inc., under cooperative agreement with the National Science Foundation.} respectively are  11.0~\AA\ and
10.1~\AA\ for \oii~$\lambda3727$ and  3.7~\AA\ and 3.3~\AA\ for \neiii~$\lambda3869$, consistent within an estimated 10\% error due to blending and continuum level.  
This sets a corresponding limit on continuum variability during the 3.3 year interval between the dates of the SDSS (2005 January 19) and \het\ spectra.

The binary black hole model of \citep{bogda08} and \citep{dotti08} predicts a secular shift of the
b-system narrow lines because of the orbital motion (see discussion below).  In this interpretation the
r-system narrow lines represent the host galaxy and should be a stable frame of reference. We have measured the  velocity difference between the peaks of the r-system and b-system narrow lines in the SDSS and \het\ spectra.
The  results in \kms\ for (\sdss, \het) are $(2565\pm30, 2575\pm20)$ for \oii~$\lambda3727,$
$(2679\pm20, 2694\pm30)$ for \neiii~$\lambda3869,$ and
$(2811\pm30, 2756\pm40)$ for \nev~$\lambda3426).$  (The errors are based on the extreme reasonable cursor positions and on a comparison of cursor settings by eye with Gaussian fits in SPLOT constrained to fit the line core and exclude the blue wing.)  
The corresponding velocity change  in
the sense $\Delta v_{rb} \equiv  v_r - v_b$ is $+10\pm36$, $+14\pm26$, and $-55\pm50$
for \oii, \neiii, and \nev, respectively.   Note that this refers to the full 3.3~year interval. 
The mean weighted by the inverse square error is
$\Delta v_{rb} = -2\pm23$ for all three lines and $\Delta v_{rb} = +12\pm25$ for \oii\ and \neiii\
only.  The annual rates of change are $d v_{rb}/d t = -1\pm7$~\kms\ per year for all three
lines and $d v_{rb}/d t = +4\pm8$~\kms\ per year for \oii\ and \neiii\ only.
Implications for the binary model are discussed below.

\section{The third redshift}
\label{sec:isystem}

Motivated by the suspected presence of a stellar Ca K line at an intermediate redshift between
$z_b$ and $z_r$ (see below), we examined the spectrum of \j0927 for possible
emission lines at such a redshift.  The SDSS spectrum (Figure \ref{fig:oiiifig}) does indeed
show a weak, narrow emission line at $\lambda8525.5\pm1.0$ 
($z=0.7028\pm0.0002$).  This feature has a flux of $(23\pm5)\times10^{-17}~\ergcms$
and an observed-frame equivalent width (EW) of $4\pm1$~\AA.
The corresponding \oiii ~$\lambda4959$ line may be present
at the expected wavelength and intensity but is obscured by noise and night sky lines.
There is also a weak, narrow emission line at $\lambda6347.0\pm1.0$ (Figure \ref{fig:caiifig})
that corresponds to
\oii~$\lambda3727.4$ at $z=0.7028\pm0.0003$.   
This feature has a flux of $(16\pm4)\times10^{-17}~\ergcms$
and an observed-frame equivalent width (EW) of $1.7\pm0.4$~\AA. 
The close agreement of the \oiii\ and \oii\
redshifts indicates the reality of the system.  We call this intermediate redshift system
the ``i-system"  and take $z_i = 0.7028$.   The \het\ spectrum confirms the presence of
the i-system \oii\ line at a velocity and intensity (EW $2.8\pm0.7$~\AA) consistent with SDSS.  
The upper limit
on any narrow  \hbeta\ emission line at $z_i$ is about one-third the flux in $\lambda5007$.
These line intensities are consistent with emission from a low luminosity AGN
or with H II region emission corresponding to a star formation rate
of about one solar mass per year \citep{kennicutt98}.

The i-system offers an alternative
candidate for the host galaxy of the QSO, or it may represent a close companion.  In either
case, the presence of a galaxy at a velocity close to the broad-line velocity lends
credence to the possibility that the r-system may not represent the host galaxy and
that the broad line system may be an ordinary QSO.  The i-system could also represent
gas ejected from the QSO.

\section{Stellar absorption features}
\label{sec:stellar}

\begin{figure}[ht]
\begin{center}
\plotone{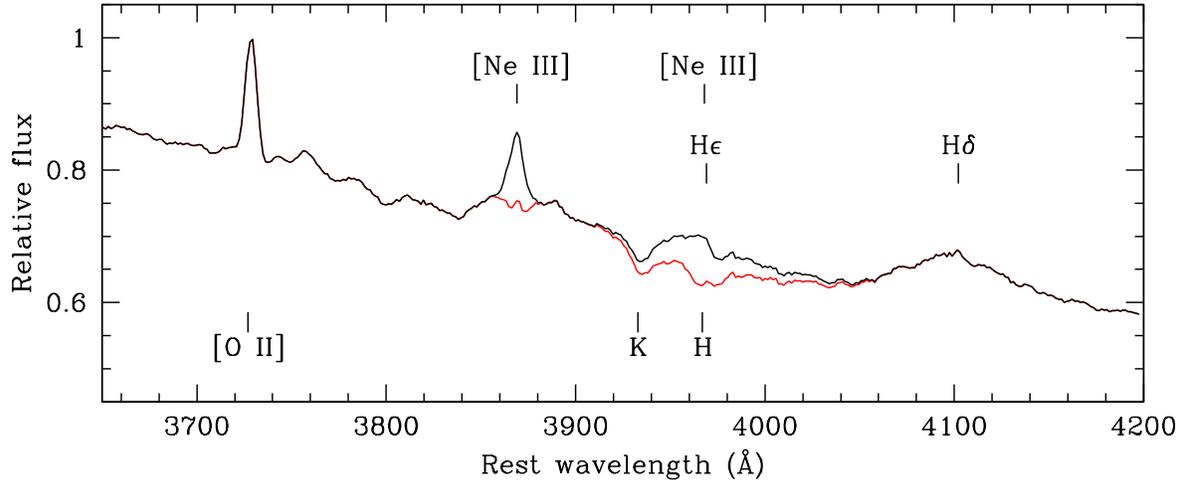}
\figcaption[]{
Composite spectrum of SDSS quasars with values of $L/\led$ similar to that of \j0927 (see text).
The region of the stellar calcium H and K lines is shown with and without subtraction 
of  \neiii~$\lambda3968$ and \hepsilon.  The calcium K line at $\lambda3933$ is clearly visible.
\label{fig:compfig} }
\end{center}
\end{figure}

The presence of stellar lines at a redshift
close to that of the broad emission lines  of \j0927\ would undermine the recoil
hypothesis.  
The stellar features from the host galaxy in QSO spectra are typically quite weak, 
particularly at higher redshifts, because of
the predominance of the AGN continuum.  The
\caii\ K line at $\lambda3933$  is often the most visible
feature \citep{greene06}.  If the black hole mass is proportional to the
host galaxy luminosity \citep[][and references therein]{lauer07}, 
the ratio of galaxy to QSO continuum
should scale inversely as the Eddington ratio $L/\led$.
From the width of the broad emission lines and the
continuum luminosity, \citet{komossa08} find 
$\mbh = 10^{8.8}~\msun$ for the black hole in \j0927,
giving $L/\led = 10^{-1.0}$.  
As a guide to the expected strength of the Ca II lines in \j0927, 
we constructed a composite spectrum  using QSOs 
from SDSS Data Release 5 (DR5) having $0.6 \leq z \leq 0.8$  and  $ -1.4  < \log{L/\led} < -0.6$,
similar to $z$ and $L/\led$ for \j0927.  The spectra were processed in the manner
described by \citet{salviander07} for their  ``HO3'' sample, shifted to the rest wavelength scale using the SDSS redshift, and scaled to a common value of $\flam$ at $\lambda4000$.
Figure \ref{fig:compfig} shows the region of the \caii\ lines in the composite spectrum for the 2181 QSOs selected in this fashion.  The \caii~K line is evident at $\lambda3933$,
with a depth of about 5\% below the total continuum.  This is consistent with the composite QSO
spectrum of \citet{vandenberk01}.

The companion line in the \caii\ doublet at $\lambda3968$ is masked by
the narrow \neiii~$\lambda3968$ and broad \hepsilon\ at $\lambda3970$ emission lines,
but an approximate subtraction is possible.
We subtracted \neiii\  using $I(\lambda3968) = 0.31 I(\lambda3869)$ 
\citep{osterbr06} and a Gauss-Hermite fit to the profile of $\lambda3869$ .   For \hepsilon, we examined the Balmer decrement
in a  generic model for a BLR cloud computed with version 07.02.00 of the photoionization code
CLOUDY,  most recently described by \citet{ferland98}.   This was a plane-parallel model
with solar abundances, gas density $N = 10^{10}~\cmmithree$, ionization parameter
$U = 10^{-2}$, the ``table power law" ionizing continuum, and a stop column density of
$10^{22}~\cmmitwo$.  The model gave
 $I(\hepsilon)/I(\hdelta) = 0.60$, compared with 0.62 for low density ``Case B''
 and 0.53 to 0.55 in several other CLOUDY models with the same density but different ionizing
 continua and larger column densities.  We
subtracted \hepsilon\ using a Gaussian profile with a wavelength-integrated flux of
$0.6 F(\hdelta)$ and the same FWHM as \hdelta.
The subtraction reveals the Ca~H line.
The depth of the K line is about 7\% in the subtracted spectrum,  
in rough agreement with the expected value  for this range of $L/\led$ on the
basis of the black hole - galaxy luminosity relationship of Lauer et al.  (2007) for bright ellipticals.

We initially suspected the presence of a weak Ca~K absorption line at a redshift similar
to $z_i$, based on visual inspection of the SDSS spectrum.  We performed a subtraction of the r- and b-system \neiii\ and \hepsilon\  lines from the SDSS spectrum of \j0927\ in the manner described above.  The resulting
spectrum (Figure \ref{fig:caiifig}) shows no convincing K line at any of the three redshifts.
The K line is not evident in the \het\ spectrum (Figure \ref{fig:hetfig}), which
has higher signal-to-noise.  
We estimate an upper limit of 10\%\ on the depth of the K-line, consistent with the expected depth as
discussed above.   Detection of host galaxy absorption lines is an important goal for future
observations.

\section{Geometrical constraints}
\label{sec:geom}

The likelihood of observing a high velocity recoil such as proposed for \j0927
depends on the length of time during which it would have the current appearance.
\citet{komossa08} suggest an upper limit of $\sim10^9$~yr for the post-kick duration of
AGN fueling, if the
mass of the bound disk is close to that of the black hole and the efficiency of
luminosity production is as high as expected for a rapidly spinning hole.  
\citet{blecha08} estimate
a bound disk mass of $\sim 10^7~\msun$, based on a disk model that
takes account of the disk's self gravity, giving a lifetime of $~\sim10^7$~yr.   The object likely
would not maintain its current appearance over even this shorter time.  
At the observed velocity, the black hole would reach
30~kpc from the galactic nucleus in $10^{7.1}$~yr, by which time
the r-system narrow lines would fade as the galactic ISM intercepted a diminishing fraction
of the ionizing continuum from the recoiling AGN.

A still tighter constraint follows from the ionization equilibrium of the gas emitting
the r-system narrow lines. The r-system
\oii~$\lambda\lambda3728.8, 3726.0$ doublet is marginally resolved in the
SDSS spectrum, with equal intensities.  
Given the \oiii\ redshift,  the observed wavelength of 
$6386.0\pm0.2$~\AA\ of the \oii\ doublet also supports equal intensities.   
For $r = I(3728.8)/I(3726.0) = 1.0 \pm 0.2$, based on the appearance of the doublet and the mean wavelength,  the ``NEBULAR''  software\footnote{http://stsdas.stsci.edu/nebular} \citep{shaw95}  gives
$N_e = 380~\cmmithree$  (range $140$ to $670~\cmmithree$).  The corresponding value
of the \sii\ doublet ratio $I(6717)/I(6730)$ is 1.12, a typical NLR value 
\citep{salviander07}.    In the recoil model, 
the ionizing source for the r-system lines is the broad line AGN.  
The observed AGN continuum flux at $\lambda5100$
rest wavelength corresponds to $\lamLlam = 10^{44.96}~\ergps$, 
giving $\lbol \approx 9\lamLlam = 10^{45.92}~\ergps$ following \citet{kaspi00}.
If we take an ionizing luminosity of $0.3\lbol$ and a mean ionizing photon
energy of 2 Ryd, then the ionizing photon luminosity is $Q = 10^{55.76}~\mathrm{s}^{-1}$,
consistent with the broad \hbeta\ luminosity of $10^{45.25}~\ergps$.  The uncertainty in $Q$ may
be $\sim0.3$~dex, based on the assumptions.  NLR models
with CLOUDY (solar abundances, density $N = 10^3~\cmmithree$, and either a $L_\nu \propto \nu^{-1}$ or
``Table AGN'' ionizing continuum) reproduce the \oii/\oiii\ line ratio of
\j0927 for an ionization parameter $U \equiv \phi/Nc = 10^{-2.5\pm0.2}$, where
 $\phi = Q/4\pi R^2$.   For $N = 380~\cmmithree$
and $Q = 10^{55.76}~\mathrm{s}^{-1}$, this
places the ionizing source at a distance of 3.5~kpc from the
gas.  This is a plausible distance in terms of allowing gas in a galactic disk to
intercept a substantial fraction of the ionizing radiation and give
the strong narrow lines of the r-system.     The derived radius depends on parameters
as $R \propto Q^{1/2} N^{-1/2} U^{-1/2}$.  This gives an uncertainty of $\pm0.23$~dex in $R$
if we add the three contributions in quadrature, or $\pm0.8$~dex if we force each uncertainty
toward the small $R$ or large $R$ extreme.

At $2650~\kms$, the
recoiling black hole requires only $10^{6.1}~\yr$ to reach 3~kpc from the
galactic nucleus.   Thus, we are catching this object at a fortuitous
moment, if it is indeed a recoil event.  A similar photoionization argument is given
by \citet{bogda08} and \citet{heckman08}.

At a distance of 3~kpc from the galactic nucleus, the projected angular separation
is $0.4\sin\theta$~arcsec.  The angle $\theta$ of the recoil velocity to the line of slight may be
small, since the radial velocity of $2650~\kms$ is close to the maximum theoretical
kick velocity of $4000~\kms$.  Thus the angular offset of the AGN from the host galaxy
nucleus may or may not be resolvable by \hst\ in the recoil picture.

\section{Discussion}

\subsection{Superposition in a Cluster?}

The possibility of chance superposition of a second AGN was rejected as improbable by \citep{komossa08}.
The SDSS image of \j0927\ appears point-like, and the SDSS spectrograph fibers have a diameter 
of 3~arcsec.  The alignment must be within $\sim1$~arcsec on the sky.   
From the overall QSO luminosity function \citep{croom05, richards06}, 
the odds of a chance superposition within 1~arcsec and within $2650~\kms$ is only $\sim 10^{-8}$
for a given QSO.  This assumes that narrow line QSOs have
similar abundance to their broad line counterparts.  However, the probability is substantially
enhanced by clustering.  
Given the existence of a broad line QSO in a cluster in the first place, what is the
likelihood of a second, superimposed narrow line AGN?
A massive cluster is needed to support the observed velocity
difference.  Given an incidence of AGN per galaxy
at this redshift of $\sim10^{-2}$  \citep[][and references therein]{shi08}, 
it may be reasonable to assume
that there typically is one additional narrow line QSO in the cluster.
For a cluster radius of
$1$~Mpc, the probability of a superposition within 1~arcsec is $~10^{-4.3}$.     
This suggests that several such superpositions may be present among the $\sim10^5$ QSOs in SDSS, or even the subset with $z < 0.8$ having \oiii\ in range.

The SDSS images
show a number of galaxies in the vicinity of \j0927\ whose photometric redshifts
\citep{csabai03}, given by the SDSS data server
\footnote{http://www.sdss.org}, are consistent with \j0927\ within the errors  of $\sim0.2$.
The SDSS photometric query server gives four galaxies of magnitude $i = 20.0$ to 22.2 within 0.5 arcminute (200~kpc).  One, with $i=21.6$ and photometric redshift $0.63\pm0.25$, is only 4.7 arcsec (33~kpc) from the QSO.   Other apparent companions within a few arcsec are visible in the images.
An apparent  cluster of galaxies lies about 0.5~arcmin to the southwest
of \j0927.  On the eastern edge of  this group, 33~arcsec SSW of the QSO, is 
SDSS J092711.93+294312.3 with  $i=20.3$, $M_i = -22.8$, and $z_\mathrm{phot} =0.66\pm0.02$,
close to the redshift of \j0927. 

Consistent with the presence of close companions is the intermediate redshift system
 that we have identified in \j0927.
The broad lines have a line-of-sight velocity of only $900~\kms$ 
with respect to this redshift, reasonable for orbital motion 
in a galactic merger or a collision or superposition
in a cluster of galaxies.  The black hole mass of $10^{8.8}~\msun$ corresponds to
a stellar velocity dispersion of $\sigstar = 300~\kms$ by the local \mbhsigstar\
relationship \citep{trem02}.   Orbital velocities of several
times \sigstar\ can occur in mergers \citep{merritt04}.  
The velocity of the r-system
narrow lines is $1700~\kms$ relative to $z_i$.  This is large for a galactic merger  but
is possible in a cluster of galaxies \citep{hayashi06}.   

While this manuscript was in preparation, a preprint by \citet{heckman08}  appeared suggesting that
\j0927\ is a high-redshift analog of the nearby active galaxy NGC 1275.  This object  has two sets
of narrow emission lines separated by $3000~\kms$ representing two interacting galaxies.   
These authors also note the apparent cluster in the vicinity of \j0927, and they suggest that
the r-system lines of \j0927\ may be gas in an interacting galaxy photoionized by the QSO continuum.

\subsection{A Recoiling Black Hole?}

\citet{komossa08} note that the b-system
narrow lines are rather broad and asymmetrical and suggest that this supports a special nature for the object as opposed to a superposition.    However, wider forbidden lines are
associated larger \mbh\ and $L$.  The b-system \oiii\ widths of $460~\kms$ \citep{komossa08}
resemble the average value of $444~\kms$ for the QSOs of \citet{salviander07} having
$\nuLnu > 10^{45}~\ergps$ and $\mbh > 10^{8.5}~\msun$, similar to \j0927.
Blue wings on \nev\  do occur at velocities  similar to \j0927 ($-1500~\kms$ from the line peak).  Our composite SDSS QSO spectrum shows such a blue wing, although not as strong as in \j0927.

How unlikely is a combination of parameters giving a line-of-sight
recoil velocity of $2650~\kms$?   As noted above, extrapolation of numerical results indicates a maximum possible kick of $4000~\kms$ for equal mass holes with maximal spins anti-aligned and lying in the orbital plane.   \citet{baker08} find that the recoil velocity varies as $\eta^3$, where
$\eta \equiv q/(1+q)^2$ and $q = m_1/m_2  < 1$ is the black hole mass ratio.  If other parameters
are optimal for a 4000~\kms\ kick, then a kick of 2650~\kms\ is possible for $q > 0.5$.  In
the cosmological merger simulations of \citet{sesana07}, merger mass ratios are
fairly uniformly distributed in log~$q$ for their ``large seeds" scenario.  If a luminous QSO typically involves a major merger with $q > 10^{-1}$, then a fraction
$\sim10^{-0.5}$ of all QSOs may have a mass ratio capable of giving the required kick.  
Spins of black holes in AGN are not well determined, but it is commonly believed that
they are often $a_* = 0.9$ or larger  \citep[e.g.,][]{fabian05}, consistent with
kicks of $\sim3000~\kms$ or larger if the other parameters are optimal \citep{baker08}.
\citet{campa07a} find that the kick velocity varies sinusoidally with
 the initial spin direction of each hole relative to its linear momentum.
Assuming a distribution in $q$ above $10^{-0.5}$  and $a_*$ above 0.9, the typical
kick for perfect spin alignment will be some average over the range $2650$ and $4000~\kms$.  
If this requires that the two spins must be anti-aligned such that $\mathrm{cos}\, \theta > 0.9$, 
and that the common axis lie in the orbital plane within a similar tolerance,
then a fraction $\sim 10^{-2}$ of randomly aligned mergers will have the needed alignment.  
If objects with a true kick between 2650 and $4000~\kms$ have a typical value $\sim 3300~\kms$,
then a fraction $10^{-0.7}$ will have a kick over $2650~\kms$
projected onto the line of sight to the observer, assuming that
redshifted or blueshifted kicks would equally attract notice.   Then
the fraction of QSOs whose line-of-sight kick exceeds $2650~\kms$ will be 
$~\sim 10^{-0.5} \times 10^{-2} \times 10^{-0.7} = 10^{-3.2}$.  We have argued that
\j0927\ is being observed in a stage of evolution lasting only $10^6$~yr after the recoil event.
If we assume that typical QSOs shine for a Salpeter time of $\sim10^{7.6}$~yr, then the
odds of catching a particular QSO in the \j0927\ stage is only $\sim10^{-1.6}$.
Altogether, this gives a probability of only $~\sim10^{-4.9}$ that a QSO will show a kick of
the magnitude of \j0927.  The SDSS DR6 has 90,000 QSOs at $z < 2.3$, of which approximately
one-fourth have $z < 0.8$ allowing observation of the \oiii\ lines.  The expected number
of QSOs with kicks over $2650~\kms$ is therefore approximately $10^{-0.4}$.  This
very approximate estimate suggests that finding one example
of a large kick in the SDSS data base is not altogether implausible.    This assumes
(1) that spin orientations are random, (2) that spin magnitudes are large, and (3) that
the final merger occurs during the QSO phase.  If some physical process tends to align the
spins, as proposed by \citet{bogda07},  then large kicks will not occur. 

\subsection{A Supermassive Black Hole Binary?}

The binary black hole model of \citet{bogda08} and \citet{dotti08} 
 assumes that the broad lines are associated with the less massive black hole of the pair, whose orbital
 motion gives rise to the blueshift of the b-system lines while the r-system reflects the host galaxy velocity.
The model has the advantage that the
binary may be in a long lived stage of evolution.  This may be as much as $10^2$ times longer than
the observable phase of the recoil scenario, before the QSO moves too far from the galaxy.
However, the binary model is constrained by our limit $dv/dt < 10~\kms$ per year on the rate of change of the velocity separation between the r- and b-systems (see \S  \ref{sec:hetspec}).   The rate of change
of the line-of-sight velocity $u_2$ of the secondary black hole $M_2$, identified with the b-system
spectral features, is given by
\citet{bogda08} as 
$du_2/dt \approx (88~\kms)(1+q)^2\,M_{2,8}^{-1}\,q_{0.1}\,s_i^{-3}\,s_\phi^{-4}\,c_\phi$,
where $q_{0.1} \equiv 10 M_2/M_1$, $M_{2,8} = M_2/(10^8~\msun)$, 
$i$ is the inclination of the orbital axis to the line-of-sight,
$s_i \equiv \mathrm{sin}~i/\mathrm{sin}~45\deg$,
$s_\phi \equiv \mathrm{sin}~\phi/\mathrm{sin}~45\deg$, 
and $c_\phi \equiv \mathrm{cos}~\phi/\mathrm{cos}~45\deg$.
There is a range of orbital phase
around $\phi = 0$ (inferior conjunction) and $\phi= 180\deg$ for which $du_2/dt$ exceeds 
the observational limit.   For some parameters, this
requires the the orbital phase to be close to quadrature where $dv/dt$ is zero ($\phi = 90\deg$
or $270\deg$).   The severity of the constraint depends on the system parameters.  For larger
black hole masses, the period is longer, the orbital acceleration is smaller, and the constraint 
on $\phi$ is less severe.  For small $i$, the orbital velocity and acceleration are large and
the constraint on $\phi$ is severe.   A modest inclination may be favored
by the unified model of AGN.   \citet{dotti08} suggest a model with
$M_1 = 6\times10^8~\msun$, $M_2 = 1.7\times10^9~\msun$, and $i = 40\deg$.
(This large $M_2$ may stretch the upper limit on the Ca K line  in \S \ref{sec:stellar}.)
For these parameters, our constraint $du_2/dt < 10~\kms$ restricts $\phi$ to be
within $\pm12\deg$ of quadrature, corresponding to 13\%\ of each orbital period.
Alternatively, \citet{bogda08} argue that in the binary model, the broad emission-line width may not
give a valid indication of the AGN black hole mass ($M_2$ in the binary model).  
From the observed X-ray luminosity,
an assumed bolometric correction, and the Eddington limit, they argue for a lower limit
of $M_2  > 5\times10^7~\msun$.  They also argue that $M_2$ should be substantially less
than $M_1$ in order for the AGN luminosity of $M_1$ to be relatively unimportant,
consistent with a single set of broad emission lines in the observed spectrum.
For $M_{2,8} = 0.5$ and $M_{1,8} = 5$, the allowed azimuth range around quadrature
is $(\pm 8, \pm3, \pm0.9\deg)$ for $i = 45, 30, 20\deg$, respectively.
We would be observing the binary in a special portion of only $\sim10^{-1}$ of
its orbital period, somewhat offsetting the advantage of the binary model as
a long lived configuration.

Other issues for the binary model, as noted by \citet{heckman08}, include the presence of
strong \oii\ emission in the b-system.  This is difficult to account for at the densities $n_e > 10^6~\cmmithree$
derived by \citet{dotti08} for the tidal gap in the accretion disk that is the proposed source of the b-system narrow lines.  Also of interest is how the binary model would give the observed, normal broad line profiles and the close redshift agreement of the b-system broad and narrow lines.

More exotic interpretations of \j0927 can be imagined.  For example, if the host galaxy
is indeed at an intermediate redshift, the system might be
a slingshot event in which a merging triple black hole system has ejected one black hole
at $+1700~\kms$ while the the remaining binary rebounded toward the observer at $-900~\kms$.   Problematic in such a picture is how the r-system object could retain or
reacquire an NLR with line widths of only $170~\kms$
after experiencing a kick of $1700~\kms$.  

\section{Conclusion}
\label{sec:conclusion}

The key question about \j0927\ is whether it is indeed a high velocity recoil or some other dramatic event.
Confirmation of such a recoil would be an important confirmation of recent
predictions of numerical relativity.  Additional observations of the candidate system are needed
to establish its true nature.   \citet{komossa08} note the importance of
obtaining \hst\ imaging to resolve the recoiling AGN from its host galaxy.
Most essential is to establish the velocity of the host galaxy; detection
of stellar features close to the b-system (broad line) velocity would suggest a normal
QSO at rest in its host galaxy.
We have argued that the \caii\ lines provide an opportunity to
do this with high quality spectra.   Other stellar features may be more difficult,
given the presence of AGN emission lines near the G-band, the Mg b-band,
and the Na D lines.  The infrared calcium triplet, seen in the composite AGN
spectrum of \citet{vandenberk01}, may be an alternative.  

Also important are
imaging and spectroscopic studies of the environment of \j0927 to establish
the nature of the galaxy cluster that apparently contains the QSO.   
If the velocities of the neighboring
galaxies resemble the b-system (broad line) redshift
of \j0927\  this will weigh against a recoil or binary model.
Finally, if the narrow lines of the r-system come from the ISM of the host galaxy,
photoionized by a displaced QSO, spectroscopic anomalies might be expected.
For example, in the geometry deduced above, the QSO radiation flux reaching
the r-system gas is similar to that in Galactic and extragalactic H II regions.
Iron remains depleted into grains in these H II regions, whereas it is gaseous
in the NLR of AGN.  The intensity of emission lines such as \fevii~$\lambda 6087$
in the r-system of \j0927\ may be a diagnostic.  Also, the \nev\ lines of AGN are
often broader than the lower ionization lines.  Confirmation of this for the
r-system lines of \j0927\ would indicate a normal AGN with the ionizing source in
the center of the NLR.  This would argue against recoil and against the variant
of the superposition model in which the r-system lines are the ISM of a passing non-active
galaxy photoionized by the main AGN \citep{heckman08}.

Pending such studies,
the existence of a a third velocity system in the spectrum of \j0927\ as noted
here, relatively close to the broad line velocity, together with the presence of a substantial cluster
apparently containing \j0927, suggests that the superposition hypothesis deserves attention.

\acknowledgments
We thank   Anita Cochran, Mike Gladders, Tesla Jeltema, Stefanie Komossa, Milos Milosavljevi\'{c}, and David Rosario for helpful discussions and assistance.
G.S. acknowledges support from the Jane and Roland Blumberg
Centennial Professorship in Astronomy at the University of Texas at Austin.

Funding for the Sloan Digital Sky Survey (SDSS) has been provided by the Alfred P. Sloan Foundation, the Participating Institutions, the National Aeronautics and Space Administration, the National Science Foundation, the U.S. Department of Energy, the Japanese Monbukagakusho, and the Max Planck Society. The SDSS Web site is http://www.sdss.org/. The SDSS is managed by the Astrophysical Research Consortium (ARC) for the Participating Institutions. The Participating Institutions are The University of Chicago, Fermilab, the Institute for Advanced Study, the Japan Participation Group, The Johns Hopkins University, the Korean Scientist Group, Los Alamos National Laboratory, the Max-Planck-Institute for Astronomy (MPIA), the Max-Planck-Institute for Astrophysics (MPA), New Mexico State University, University of Pittsburgh, University of Portsmouth, Princeton University, the United States Naval Observatory, and the University of Washington.

The Hobby-Eberly Telescope (HET) is a joint project of the University of Texas at Austin, the Pennsylvania State University, Stanford University, Ludwig-Maximilians-UniversitŠt MŸnchen, and Georg-August-UniversitŠt Gšttingen. The HET is named in honor of its principal benefactors, William P. Hobby and Robert E. Eberly.  The Marcario Low Resolution Spectrograph is named for Mike Marcario of High Lonesome Optics who fabricated several optics for the instrument but died before its completion. The LRS is a joint project of the Hobby-Eberly Telescope partnership and the Instituto de Astronom'a de la Universidad Nacional Aut—noma de MŽxico.

%Table here for mss.

\clearpage

\clearpage

\clearpage

\clearpage

\clearpage


\begin{thebibliography}{38}
\expandafter\ifx\csname natexlab\endcsname\relax\def\natexlab#1{#1}\fi


  \bibitem[{Baker} {et~al.}(2008)]{baker08}
{Baker}, J.~G., {Boggs}, W.~D., {Centrella}, J., {Kelly}, B.~J., {McWilliams},
  S.~T., {Miller}, M.~C., \& {van Meter}, J.~R. 2008, \apjl, 682, L29.

\bibitem[{{Begelman} {et~al.}(1980){Begelman}, {Blandford}, \&
  {Rees}}]{begelman80}
{Begelman}, M.~C., {Blandford}, R.~D., \& {Rees}, M.~J. 1980, \nat, 287, 307
  
 \bibitem[Blecha \& Loeb(2008)]{blecha08} Blecha, L., \& Loeb, A. 2008, \mnras, 390, 1311
 
\bibitem[Bogdanovi{\'c} et al.(2007)]{bogda07} Bogdanovi{\'c},
T., Reynolds, C.~S., \& Miller, M.~C.\ 2007, \apjl, 661, L147
 
\bibitem[Bogdanovi{\'c} et al.(2008)]{bogda08} Bogdanovi{\'c},
Eracleous, M., \& Sigurdsson, S. 2008, arXiv:0809:3262
  
\bibitem[{{Bonning} {et~al.}(2007){Bonning}, {Shields}, \&
  {Salviander}}]{bonning07}
{Bonning}, E.~W., {Shields}, G.~A., \& {Salviander}, S. 2007, \apjl, 666, L13

\bibitem[{{Campanelli} {et~al.}(2007{\natexlab{a}}){Campanelli}, {Lousto},
  {Zlochower}, \& {Merritt}}]{campa07a}
{Campanelli}, M., {Lousto}, C., {Zlochower}, Y., \& {Merritt}, D.
  2007{\natexlab{a}}, \apjl, 659, L5  

\bibitem[{{Campanelli} {et~al.}(2007b){Campanelli}, {Lousto}, {Zlochower}, \&
  {Merritt}}]{campa07b}
{Campanelli}, M., {Lousto}, C., {Zlochower}, Y., \& {Merritt}, D. 2007, Phys. Rev. Lett.,
 98, 231102
 
 \bibitem[Croom et al.(2005)]{croom05} Croom, S. M.,  Smith, R. J.,  Boyle, J., Shanks, T., 
  Miller, L, Outram, P. J., Loaring, N. S., Hoyle, F., \& da \^Angela, J.  2005,  \mnras, 349, 1397
     
\bibitem[{{Csabai} {et~al.}(2003){Csabai} et al.}]{csabai03}
Csabai, I., et al. 2003, \aj, 125, 580

 \bibitem[Dain et al.(2008)]{dain08} Dain, S., Lousto, C. O., and Zlochwer, Y. 2008,  
     Phys. Rev. D, 78, 024039
 
  \bibitem[Dotti et al.(2008)]{dotti08} Dotti, M., Montuori, C., Volonteri, M., Colpi, M, \& Haardt, F. 
       2008,  arXiv:0803:3446

\bibitem[Fabian(2005)]{fabian05} Fabian, A. C. 2005, \apss, 300, 97

\bibitem[Ferland et al.(1998)]{ferland98}
Ferland, G. J., Korista, K.T., Verner, D.A., Ferguson, J.W., Kingdon, J.B. \&
Verner, E.M. 1998, PASP, 110, 761 


\bibitem[Greene \& Ho(2006)]{greene06}  Greene, J. E., \& Ho, L. 2006, \apj,  641, 117

\bibitem[Hayashi \& White(2006)]{hayashi06} Hayashi, E. \& White, S. D. M.  2006, \mnras, 370, L38

\bibitem[Heckman et al.(2008)]{heckman08} Heckman, T.~M., Krolik, J.~H., Moran, S., Schnittman, J.,
  \& Suvi, G. 2008, arXiv:0810.1244

\bibitem[Kaspi et al.(2000)]{kaspi00} Kaspi, S. \etal\ 2000, \apj, 533, 631

\bibitem[Kennicutt(1998)]{kennicutt98} Kennicutt, R. C. 1998, \araa, 36, 189

\bibitem[{{Komossa} {et~al.}(2008){Komossa}, {Zhou}, \&
  {Lu}}]{komossa08}
{Komossa}, S., {Zhou}, H. \& {Lu}, H. 2008, \apjl, 678, L81

\bibitem[Komossa \& Merritt(2008)]{komerritt08}
{Komossa}, S., \&  Merritt, D. 2008, \apj, 683, L21

\bibitem[Lauer et al.(2007)]{lauer07} Lauer, T.~R., et al.\ 2007a, \apj, 662, 808

\bibitem[{{Libeskind} {et~al.}(2006){Libeskind}, {Cole}, {Frenk}, \&
  {Helly}}]{libeskind06}
{Libeskind}, N.~I., {Cole}, S., {Frenk}, C.~S., \& {Helly}, J.~C. 2006, \mnras,
  368, 1381
  
\bibitem[Lippai et al.(2008)]{lippai08}  Lippai, A., Frei, Z., \& Haiman, Z.  2008 \apjl, 676, L5

\bibitem[{{Loeb}(2007)}]{loeb07} {Loeb}, A. 2007, Phys. Rev. Lett., 99, 041103

\bibitem[Madau \& Quataert(2004)]{madau04} Madau, P., \& Quataert, E. 2004,  \apj, 606, L17

\bibitem[{{Merritt} {et~al.}(2004){Merritt}, {Milosavljevi{\'c}}, {Favata},
  {Hughes}, \& {Holz}}]{merritt04}
{Merritt}, D., {Milosavljevi{\'c}}, M., {Favata}, M., {Hughes}, S.~A., \&
  {Holz}, D.~E. 2004, \apjl, 607, L9

\bibitem[Osterbrock \& Ferland (2006)]{osterbr06} Osterbrock, D.~E., \& Ferland 2006, 
`Astrophysics of Gaseous Nebulae and Active Galactic Nuclei,'  2nd ed., University Science Books 

\bibitem[Richards et al.(2006)]{richards06} Richards, G. T. 2006, \aj, 131, 2766

\bibitem[{{Salviander} {et~al.}(2007){Salviander}, {Shields}, {Gebhardt}, \&
  {Bonning}}]{salviander07}
{Salviander}, S., {Shields}, G.~A., {Gebhardt}, K., \& {Bonning}, E.~W. 2007,
  \apj, 662, 131
  
\bibitem[Schnittman \& Krolik(2008)]{schnitt08} Schnittman,
J.~D., \& Krolik, J.  2008, \apj, 684, 835

\bibitem[{{Sesana} {et~al.}(2007){Sesana}, {Volonteri}, \& {Haardt}}]{sesana07}
{Sesana}, A., {Volonteri}, M. \&{Haardt}, F. 2007, \mnras, 377, 1711

\bibitem[Shaw \& Dufour(1995)]{shaw95} Shaw, R. A.  \& Dufour, R.~J.  1995, \pasp, 107, 896

\bibitem[Shi et al.(2008)]{shi08} Shi, Y., Rieke, G., Donley, J., Cooper, M., 
  Willmer, C., \& Kirby, E.  2008,  \apj, 688, 794

\bibitem[Shields \& Bonning(2008)]{shields08}  Shields, G. A., \&  Bonning, E. W.,
2008, \apj, 682, 758

\bibitem[Tremaine et al.(2002)]{trem02} Tremaine, S., et al.
2002, \apj, 574, 740

\bibitem[Vanden Berk et al.(2001)]{vandenberk01} Vanden Berk, D., et al. 2001
  \aj, 122, 549

\end{thebibliography}
\end{document}